\documentclass[preprint,authoryear]{elsarticle}
\usepackage[american]{babel}
\usepackage[latin1]{inputenc}
\usepackage{amsmath,amssymb,epsfig,natbib,bm,rotating,multirow}
\usepackage[a4paper, text={16cm,21cm},top=2.5cm, left=2.5cm, right=2.5cm, includeheadfoot, headheight=1cm]{geometry}

\usepackage[T1]{fontenc}
\usepackage{float}
\usepackage{url}
\usepackage{ulem}
\usepackage{multicol}
\floatstyle{ruled}
\newfloat{algorithm}{tbp}{loa}
\providecommand{\algorithmname}{Algorithm}
\floatname{algorithm}{\protect\algorithmname}

\journal{~}

\title{MWStat: A Modulated Web-Based Statistical System}
\author{Francisco Louzada$^a$ and Anderson Ara$^a$  
 \ \\ 
{\small{$^a$ Universidade de São Paulo, Instituto de Matem\' {a}tica e Ci\^ {e}ncias da Computa\c{c}ão,  São Carlos, SP, Brazil }}   
             } 

\date{}

\begin{document}

\begin{abstract}
In this paper we present the development of a modulated web based  statistical system, hereafter MWStat, which shifts the statistical paradigm of analyzing data into a real time structure. 
 The MWStat system is useful for both online storage data and questionnaires analysis, as well as to provide real time disposal of results from analysis related to several statistical methodologies in a customizable fashion. 
Overall, it can be seem as a useful technical solution that can be applied to a large range of statistical applications, which needs of a scheme of devolution of real time results, accessible to anyone with internet access.
We display here the step-by-step instructions for implementing the system. The structure is accessible, built with an easily interpretable language and it can be strategically applied to online statistical applications. We rely on the relationship of several free languages, namely, PHP, R, MySQL database and an Apache HTTP server, and on the use of software tools such as phpMyAdmin. 
We expose three didactical examples of the MWStat system on institutional evaluation, statistical quality control and multivariate analysis. The methodology is also illustrated in a real example on institutional evaluation.
\\
\vspace{0.5cm}

\end{abstract}
\begin{keyword}
\vspace{5mm} Online Survey, Statistical  System, Real Time Results, Statistical Methods.
\end{keyword}

\maketitle

\section{Introduction}

For decades the pattern of rationality based on statistical analysis has experienced  slowness, which begins with the data collection, passes by the statistical analysis itself, and ends up in the final presentation of the results obtained in form of a report. For instance, in traditional data collect research there is a gap in time between the application of a questionnaire, the obtainment of the answers of respondents, the application of statistical techniques and the visualization of informative
reports on the final results, i.e., possibly a slow process that often
may involve a high cost. On the other hand, on account of internet
democratization, online surveys are becoming widely used. In an informal
survey over the network, \citet{kaye} identified over 2000 surveys
in 59 different areas that benefit from the technology focused on
online questionnaires. \citet{solomon}, \citet{Scott:05} and \citet{CoxCox}, amongst others, show internet based surveys offer
significant advantages over traditional survey techniques. However,
despite the wide use of online research, the gap in time between the
begin of a survey and the final presentation of the statistical
analysis results remains. 
Thus, there is a real need for a environment which facilitates
the application of a survey while directly connecting the responses in a
customizable real time informative report in order to quickly provide
all the information generated by the survey.

Over the world wide web we can find some environments that organize national statistics datasets, such as the Virtual Statistical environment housed on \url{http://www. virtualstatisticalenvironment.org/}, as well as, there are environments which collect online data focusing on creating and publishing online surveys, such as the SurveyMonkey (\url{http://www.surveymonkey.com/}) and QuestionPro (\url{http://www.questionpro.com/}). 
It is also easily found statistical systems for specific statistical analysis, such as \citet{Tung:1991}, \citet{Krtolica:1991}, \citet{Hatanaka}, Analytica  (\url{http://www.lumina.com/}), and Plug\&Score (\url{http://plug-n-score.com}), just to name a few.

Moreover, some online statistical  environments are capable to produce some simple  plots, frequency tables, cross-tabulation and summary measures. SOCR \citep{Dinov:06} (\url{http://www.socr.ucla.edu/}) proposes a suite of Java applets for statistical online computation, visualization, analysis and virtual experimentation, EasyCalculation environment (\url{http://easycalculation.com/}) is a free online math website which helps users to learn mathematics and statistics.  The above environments cannot carry out customized statistical analysis which may involve more
sophisticated procedures, such as a regression analysis, a multivariate analysis, quality control charts,
among others.  

Other environments such as the R-PHP \citep{Mineo:06} (\url{http://dssm.unipa.it/R-php/}), RStudio \citep{RStudio} and rApache \citep{rApache}  allow the utilization of $\mathtt{R}$ software directly online, that may be installed on any server. These environments have a sophisticated communication between the $\mathtt{R}$ software and servers, but do not interact dynamically with users who do not have knowledge of the $\mathtt{R}$ language. 
The environment JStatCom \citep{kratzig:07} defines some classes to connect existing math libraries (Ox, Matlab or $\mathtt{R}$ programming languages) with Java client. The Rweb environment  \citep{Banfield:99} has three different versions, the first allows to type the code, click the submit button and a page with the results is returned, the second is based on Javascript procedures and the third is designed as a point click interface that can be used in introductory statistic courses. Also, Online Analytical Statistical Information System \citep{oasis} is a online system that provides some sophisticated and public analysis of health and social science data. And, R-fiddle \citep{rfiddle} provides a free environment to write, run and share R-code right inside the browser.

Another approach for interactivity is performed by Shiny \citep{chang2015shiny},  a $\mathtt{R}$ package with a straight connection from $\mathtt{R}$ to a webserver. It is divided in two components: a user-interface script and a server script. In this case Shiny has its own structure to perform web applications, but with restricted HTML or PHP customisation.

In the present paper we built the MWStat virtual environment, with easy HTML or PHP customisation, overcoming the problems discussed above. The MWStat is based on a scheme of devolution of real time results, accessible to anyone with internet access.
In other words, these procedures are focused on how to build a user-friendly interface and how to relate any website with the $\mathtt{R}$ software to generate dynamic results for any online purposes. The procedures exposed in this paper may be easily applied by statisticians with a basic knowledge of web programming. The structure is  accessible and built with an easy interpretation  language (PHP) and strategically applied in online applications. This procedure can be considered an  system, since it replaces the manual collecting data and is able to expose more targeted results to real problems.

The MWStat relies on the relationship of several free open-source languages,
namely, PHP (\url{http//: www.php.net}), $\mathtt{R}$ (\url{http//:cran.r-project.org}),
MySQL database (\url{http//: www.mysql.com}) and an Apache HTTP server
(\url{http//:www.apache.org}), the latter for hosting and interpreting
other languages. Moreover, we used the phpMyAdmin interface (\url{http//:www.phpmyadmin.net})
as an auxiliary tool, related to the administration of MySQL database
using PHP language. These technologies were gathered in a LAMP server
(Linux, Apache, MySQL and PHP), which is a popular solution of free
open source software to build a viable web server of general purpose
with a low costing structure and high performance \citep{Neiderauer}.

Combining all the technologies above,  the MWStat can perform online collection and data analysis, and the obtained results are provided in real time, depending only on the processing time of the statistical analysis to be considered. 
Moreover, the MWStat is also very flexible, since it is completely customizable and it can be
built into independent modules, according to the user needs.

The main objective of the present paper is to present the  MWStat and the softwares involved in its construction, displaying some technical procedures on how to building it. 
Following this paper, anyone with some computational knowledge may build a web based statistical  system environment that performs online statistical analysis  for any purposes.
The versions of the software used in this work are provided, but the same procedures can be extended to different
versions or even other operating environment platform and database software.
We provide the basic codes for the implementation of the $\mathtt{R}$ environment on any LAMP server, but more detailed codes are available in our homepage (\url{http://www.mwstat.com}). 

In Section~\ref{sec:2} the softwares used for building the MWStat and their considered versions
are displayed. In Section~\ref{sec:Implementation-procedures} we
show the procedures for server installation and setup, which are
necessary for the interpretation of the languages. In Section~\ref{sec:Applications}
we present  examples of the MWStat  implementation  for  event evaluation,
as well as for two more areas,  statistical quality control and multivariate
analysis, those can be accessed in the  MWStat homepage through login and password provided.
Moreover, a fourth example is also provided on institutional evaluation, which has been used in several applied researchers in Brazil. 
We finish the paper with some final comments in Section~\ref{sec:Final-Comments}, where we also
present the  web based statistical system homepage and the various applied research developments based on its environment.


\section{Softwares applied for building the MWStat environment}

\label{sec:2}

In this section, we present the software used in the construction
of the MWStat. Essentially, they are free softwares which can be
easily found on the web. 
\begin{itemize}
\item \textbf{\footnotesize{Linux Ubuntu Server 9.10 (or higher):}} Ubuntu is a complete Linux operating
environment, completely free, with great practicality, configuration and
use \citep{Tarng}. The installation file of this operating environment
is available on website \url{http://www.ubuntu.com}, where the image
of its installation CD can be downloaded. 
\item \textbf{\footnotesize{$\mathtt{R}$ software:}} computing environment for the performance of
statistical analysis and graph building. It compiles and runs on a
wide variety of UNIX platforms, Windows and MacOS. In our case, we
used the $\mathtt{R}$ implemented in UNIX through a Linux server. 
\item \textbf{\footnotesize{MYSQL 5.0 (or higher):}} database management environment (DBMS) that uses the
SQL (Structured Query Language) as interface. It is currently one
of the most popular databases, with over 100 million installations
worldwide. 
\item \textbf{\footnotesize{PHP 5.0 (or higher):}} \citep{james} acronym for Hypertext Preprocessor,
is a language for creating script directly into the server designed
specifically for the web. Within an HTML page, PHP codes can be executed
every time the page is visited. This code is interpreted on the web
server and generates HTML viewing or other display type. Below are
listed some advantages of PHP: high performance; interfaces for many
different database environments; integrated libraries for many common tasks
from the web; Low cost, Easy to learn and use, portability, availability
to source code. Version 5.0 was developed to improve to Object Oriented
Programming and is available now in version 5.2.13 (\url{http://www.php.net}). 
\item \textbf{\footnotesize{phpMyAdmin 2.7.0-pl2: (or higher):}} computer program developed in PHP to
administer MySQL over the Internet. From this environment you can create
and remove databases; create, delete and modify tables; insert, delete
and edit fields, execute SQL code and manipulate key fields. For this
paper, we used implemented features and bug fixes up to version 2.6.2 (\url{http://www.phpmyadmin.net}). 
\end{itemize}

\begin{figure}[htb]
\begin{centering}
\includegraphics[scale=0.9]{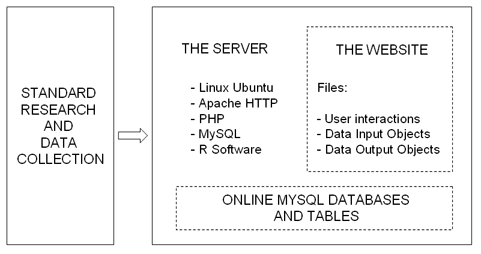} 
\par\end{centering}
\caption{General structure of the systemenvironment.\label{fig:struc}}
\end{figure}

\section{Implementation procedures}

\label{sec:Implementation-procedures}

For describing the implementation procedures, we focus 
on listing the procedures for creating the MWStat environment, as well as
the software installation, settings and programming, at the expense of  displaying the best hardware configuration
for a computer (server) to implement the system,  which is out of the scope of the paper.

We show the implementation of the environment in an Ubuntu
Linux server, with web support (Apache), PHP and MySQL, also known
as LAMP server. The Linux operating environment was installed on a specific
computer, following these steps: 
\begin{enumerate}
\item Through the installation file from the Ubuntu 9.10 Linux, we create
a CD to start the computer from it, in other words, do the boot from
the CD. If the CD does not run automatically, we must configure the
Setup (BIOS) for boot priority. 
\item Choose the preferred language (shortcut to select language: F2). 
\item Start the installation, click Install Ubuntu Server. 
\item After the detection of the network cards on the computer, we must
perform the configuration of the static IP, in other words, the number
for the network interface card eth0. 
\item Name the server: OnlineSytem. 
\item Choose a location: for example, Sao Paulo, Brazil. 
\item After that, we started the disk partitioner through the "assisted
partitioning - use entire disk" where in this case,
all necessary partitions will be created automatically. 
\item Inform the administrator user name and password. 
\item Inform the Proxy server address, if any. 
\item We must select the packages to be installed along with the environment:
the LAMP Server. 
\item Once the environment and the chosen services are installed, the server
will restart. 
\end{enumerate}

After performing the procedures above, the server created can be accessed
from any browser (eg Internet Explorer, Mozilla, Opera etc) through
its IP address, showed in Step 4.
Figure~\ref{fig:struc} shows the overall MWStat environment server structure.

We may configure the server directly on the machine installed, by
Linux' Shell. Therefore, we are interested in creating a specific
place for environment files, making it available through the Internet,
installing the $\mathtt{R}$ software and creating an environment where it can
generate graphs and communicate with the MySQL database software.

We start the configuration process by creating a user and their respective
folder, where the environment files will be. By taking the user name \texttt{usertest},
use the following command:

\medskip
\texttt{sudo useradd usertest}
\medskip

In this case, a folder is created in home\texttt{ /usertest/} where
the files created by this user will remain, we shall use this folder
to save the environment files. Likewise, we can redirect this folder to
a fixed address on the Internet. Thus, we edit the alias file with
the command:

\medskip
\texttt{sudo nano /etc/apache2/conf.d/alias}
\medskip

Notice that in this case we use the \texttt{nano} text editor, which
is a traditional editor of the Linux Ubuntu. After opening this file, insert the following lines:

\medskip
\texttt{Alias /dados/home/usertest/} 

\texttt{<Directory /home/usertest/>} 

\texttt{Options Indexes FollowSymLinks} 

\texttt{AllowOverride All} 

\texttt{Order allow,deny} 

\texttt{Allow from all} 

\texttt{</Directory>} 
\medskip

After running the commands above, files that are in the home directory
\texttt{/usertest/ }can be accessed directly through a browser using
the address \texttt{IP/data/}.

Inside the directory \texttt{/usertest/ }we can create a new directory
and save the phpmyadmin files there, where we can manage the tables
in MySQL.

In this context, we must install the $\mathtt{R}$ software on our server. We
must enter the address \url{http://cran.r-project.org/bin/linux/ubuntu/}
and after that, perform the installation. The entire procedure can
be done through the following commands:

\medskip
\texttt{sudo nano /etc/apt/sources.list}

\texttt{sudo apt-get update}

\texttt{sudo apt-get install r-base}
\medskip

After the construction to this entire structure, we still have to
provide the $\mathtt{R}$ software with the ability to generate graphics in the
Linux environment. We applied ghostscript technology and used the
following command line:

\medskip
\texttt{sudo apt-get install gs}
\medskip

We must also provide the $\mathtt{R}$ software with the ability to connect with
the MySQL database. For such, we used the package RMySQL \citep{RMYSQL},
installed with the command:

\medskip
\texttt{sudo apt-get install r-cran-rmysql}
\medskip

Directly from the server, we can run the software just by typing R
in any directory. The commands in $\mathtt{R}$ language to access the MySQL database
are displayed below (they must necessarily be in .r format and saved
on the server):

\medskip
\texttt{require(RMysQL)}

\texttt{con<-dbConnect(dbDriver("MySQL",)}

\texttt{user="username", dbname = "databasename", password="passwd")} 

\texttt{dados=dbGetQuery(con,paste("select * from tablename"))}
\medskip

After all the procedures above, we have the structure required for
the relationship among the languages in focus. Thus, we can edit \texttt{.php}
files using the resources installed on the server.

By programming in PHP, we can run directly from web pages files written
in $\mathtt{R}$ language. The codes below access files \texttt{.r} through the
PHP language,\texttt{ statistical.r} . This file refers to calculations
that have value vectors or images as output. For example, a vector
of means, which will be assigned to the variable php \texttt{\$res}
or graphs generation codes.

$\mathtt{ \$command = "echo\:  '\: argv\: <-\: \backslash"statistical.r\backslash";  }$

$\mathtt{ source(argv)\: ' "\: |\: "\: .\:  "/usr/bin/R\: \backslash \backslash\: --vanilla\: --slave"};$

\texttt{\$res = exec(\$command);}

\begin{figure*}[h]
\centering{} \includegraphics[scale=0.6]{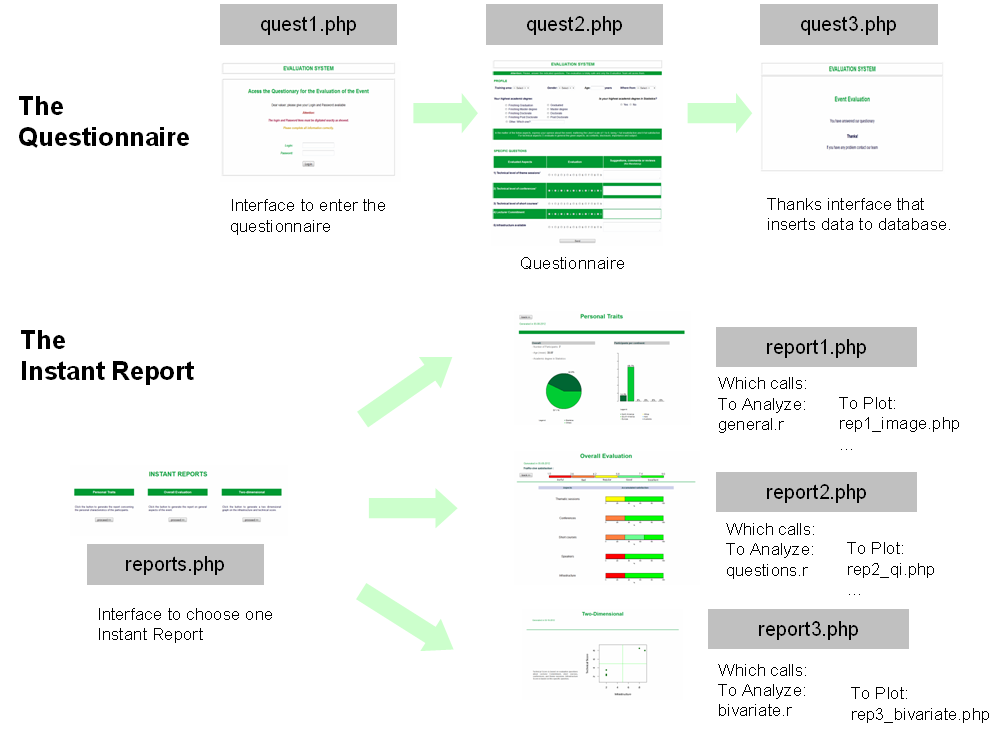}\caption{The files and their relationship in the event evaluation Example. \label{fig:sdie}}
\end{figure*}

\section{Some Applications}

\label{sec:Applications}

In this section, we display three general applications of our system.
The first, second and third examples of  applications of the system for event evaluation, quality control and multivariate analysis, respectively. They can be accessed online by the readers. Moreover, the first and third o applications can be totally reproduced in a standalone fashion by downloading the codes from \url{http://www.mwstat.com}.
The fourth application is an institutional evaluation which was carried out on an undergraduate curse in Statistics from the Universidade Federal de São Carlos, Brazil. 

\subsection{Some introductory examples}

\begin{figure}[ht]
\centering{} \includegraphics[scale=0.6]{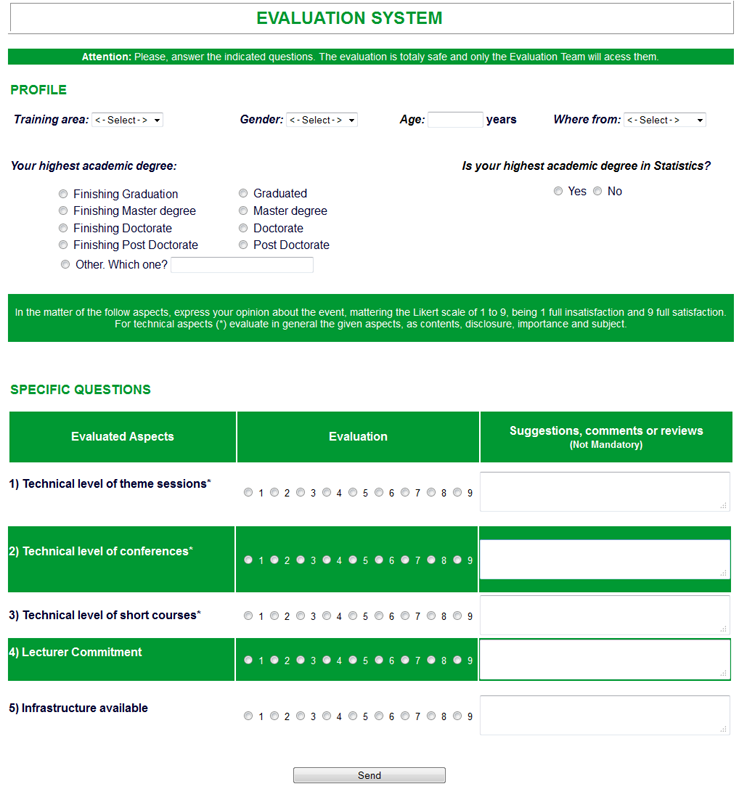} \caption{Online questionnaire for the event evaluation  example. \label{fig:quest_event}}
\end{figure}

\begin{figure}[htb]
\centering{} \includegraphics[scale=0.6]{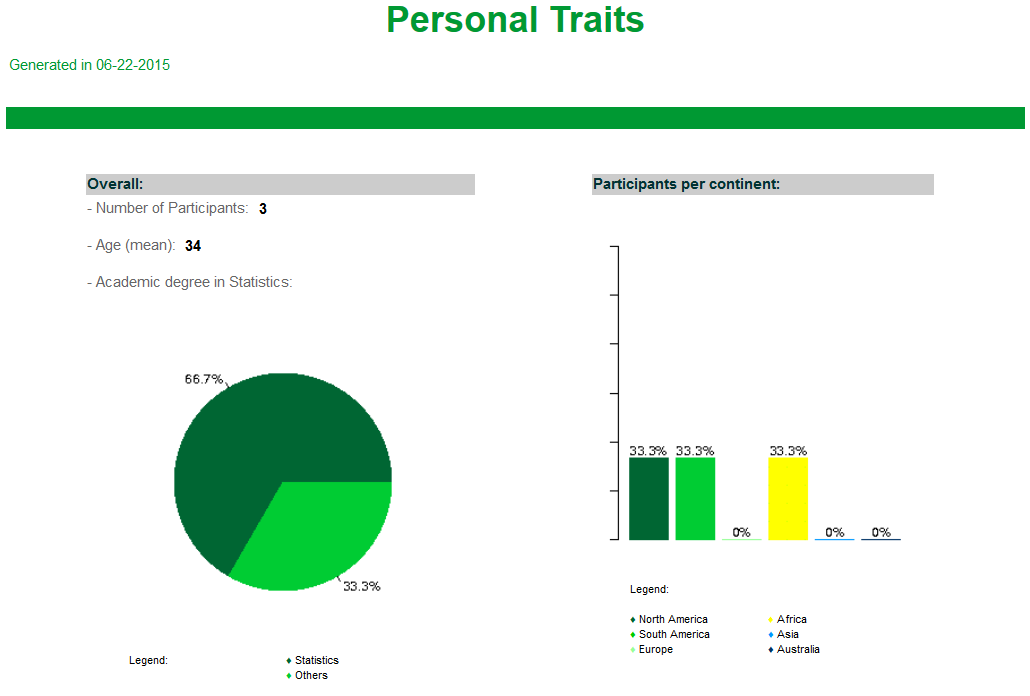} \caption{A part of the Instant personal traits report for the event evaluation  example. \label{fig:quest_rel1}}
\end{figure}

\begin{figure}[htb]
\centering{} \includegraphics[scale=0.6]{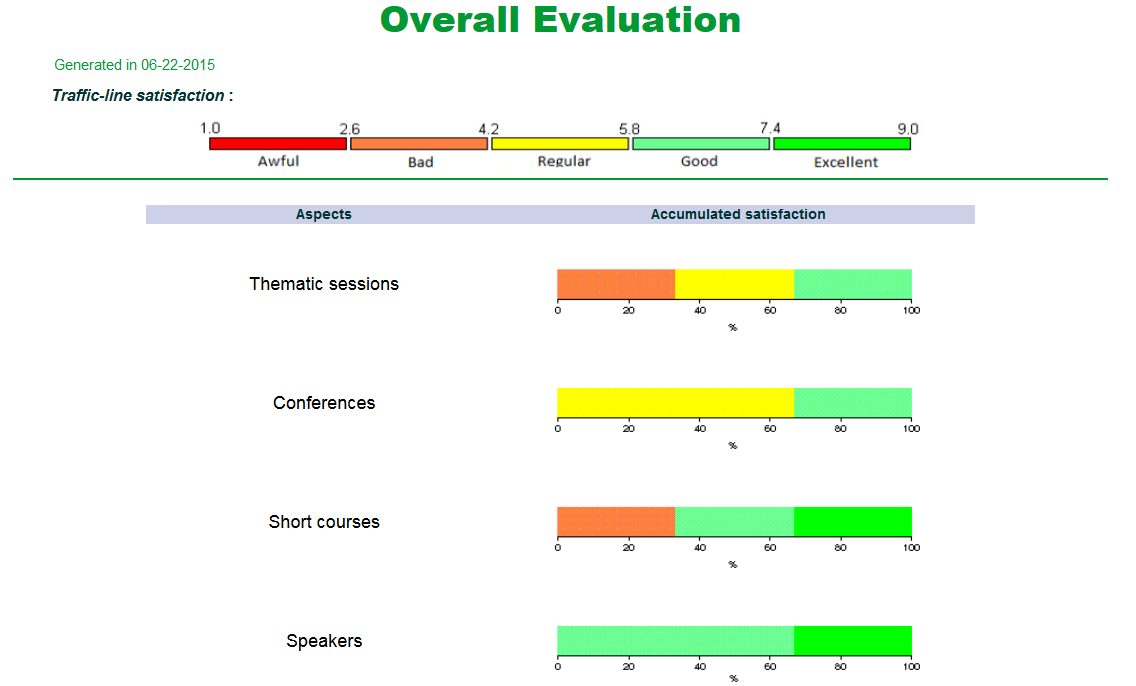} \caption{Another part of the Instant personal traits report for the event evaluation  example. \label{fig:quest_rel2}}
\end{figure}

In this section, we display more three general didactical applications of the MWStat.
The first one is based on a dynamic event evaluation which allows to measure the quality of scientific events, such as, conferences, symposia, meetings, workshops, among other. This example  consists in 11 questions about academic purposes in a generic event. Participants answer an online questionnaire and an instant online report  is provided in real time. 
The files and their relationship for this example are shown in Figure~\ref{fig:sdie}. The files may be downloaded from \url{http://www.mwstat.com} in a standalone fashion, in the sense that an interested reader can easily reproduce the overall example.
The Figure~\ref{fig:quest_event} shows the online questionnaire and Figures~\ref{fig:quest_rel1} and ~\ref{fig:quest_rel2} show a part of the instant report.

\begin{figure}[htb]
\begin{centering}
\includegraphics[scale=0.6]{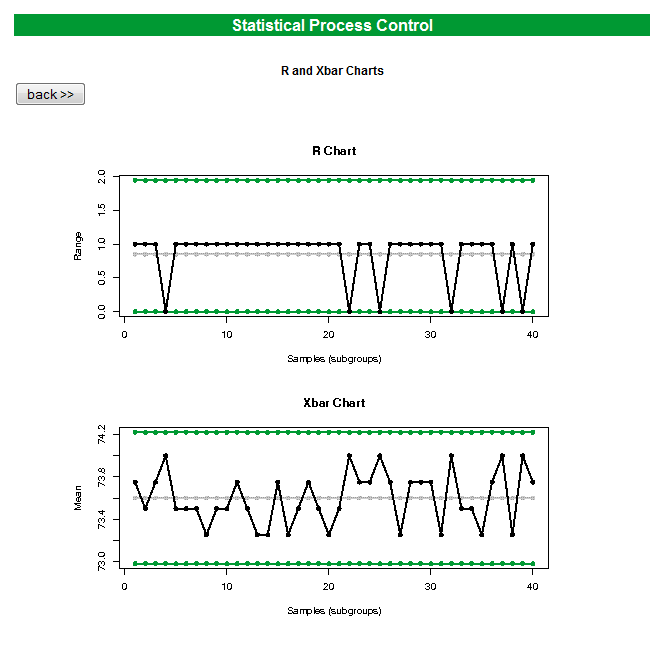} 
\par\end{centering}
\caption{A online report of the  statistical control process example. \label{fig:SCP EX}}
\end{figure}

The second application is focused in statistical process control, the default dataset was taken from Montgomery's book \citep{montgomery}. It consists of 40 samples of size 4 to control of the diameter of the piston rings. The Figure~\ref{fig:SCP EX} shows the online reports with a default dataset in SPC analysis.

\begin{figure}[htb]
\centering{} \includegraphics[scale=0.60]{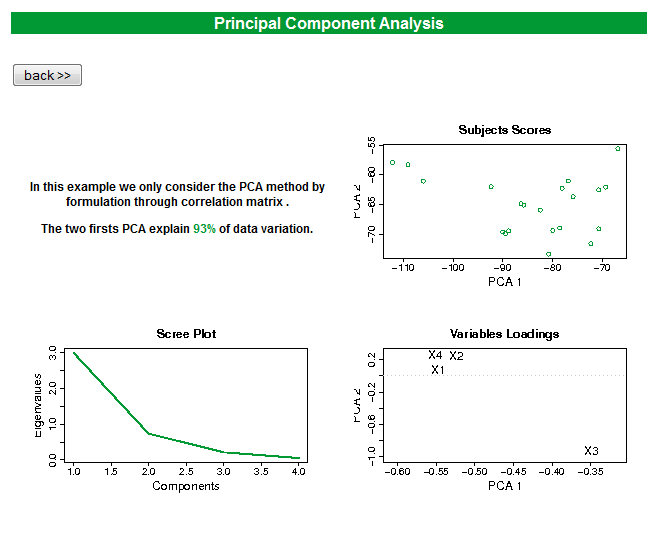} \caption{A online report of the principal component analysis  example. \label{fig:PCA EX}}
\end{figure}

The third example is based on a principal component analysis, the default dataset \citep{drapper} is composed by a sample of 4 variables about 21 days of operation of a plant oxidizing. 
The Figure~\ref{fig:PCA EX} shows the online report with a default dataset for PCA analysis.  
The files may be downloaded from \url{http://www.mwstat.com} in a standalone fashion, in the sense that an interested reader can easily reproduce the overall example.

All examples present above can be accessed online from the address \url{http://www. mwstat.com} in the article examples link. To access the examples enter the website  and provide the password \texttt{HJT534}; the uppercase letters must be maintained.
In the event evaluation example, the readers may answer the online questionnaire and access the entire instant report. The password to access this questionnaire is \texttt{mwstat}. According to the present exemplification, only one user was established, but an indefinite number of users could be set up.
In the other examples the readers may enter their own dataset or use a given default dataset available in the environment. 

\begin{figure*}[b]
\begin{centering}
\includegraphics[scale=0.65]{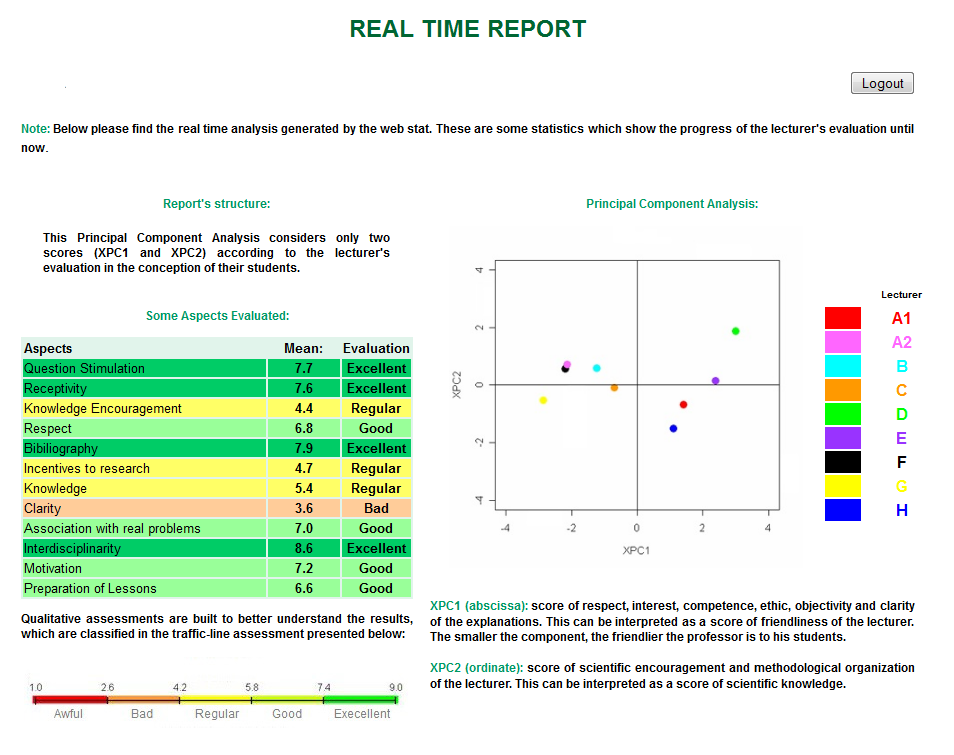} 
\par\end{centering}
\caption{Online report on view at course coordinator.\label{fig:Online-report-on}}
\end{figure*} 

\subsection{Institutional Evaluation Example}

The fourth example is related to an institutional evaluation. The main purpose of institutional evaluation is to support the university's commitments to academic and operational excellence through the collection, analyses and reporting of diversified data. The MWStat environment was used with the purpose of evaluating lectures and general aspects of a particular department within a Brazilian university. The environment was tested on a stratified sample of 165 students of an undergraduate statistics course. 
Thus, each randomly chosen student received an email containing an explanatory text followed by the website address and a random password, with which it was possible to access the restricted area designated to the students. The password was randomly generated and encrypted in the database.

After each student answered the questionnaires designated to them, the environment sent a new e-mail thanking and confirming the successful
storage of the responses. The answers are properly inserted into a MySQL database that safely only the database administrator (researcher) has access.

After each student performed his responses successfully, the homepage of the website is updated and a link to the evaluation results is
provided. An instantaneous report can be viewed throughout all the evaluation.

In the authentication area, programmed in PHP, the login and password of the individual are verified and, furthermore, the level of hierarchy of information compatible with his status is identified. In other words, how much information is provided to him. For instance, students may have a different level of information in comparison with the lectures they evaluated. In terms of illustration, let us consider that the login was made by the director for undergraduate studies in statistics. The director has access to the overall online report of his departmental lectures evaluation. An illustration of such report is shown in Figure~\ref{fig:Online-report-on}. The report displays a principal component analysis, and the general evaluation of each particular aspect. However, other statistical methodologies can be straightforwardly considered.

For sake of comparison we tried, without success, to consider the most known online statistical environments with the purpose of evaluating lectures and general aspects of the particular problem treated in this section.
Survey Monkey and QuestionPro have some free licenses able to perform one survey that contains at most ten questions and one hundred  responses, more advanced surveys can be carried out through payment; remembering 165 students were sampled.
Besides, their statistical analysis are based only on descriptive statistics and basic graphs; a principal component analysis could not be considered. SOCR has many possible statistical analyses but the environment is restricted by the developer's Java applets and it is not customizable. R-PHP does not have any online procedure enabling data collection and analysis in real time.

Thus, each in its turn is overcome by the MWStat.
Through the steps exposed in this paper, the MWStat arises as a environment which is able to carry out any number of online surveys, with unlimited number of questions and responses. Besides the MWStat performs any customized statistical analysis that may involve more general sophisticated procedures. With our open source technology it is possible to perform any kind of data collection and analysis in virtual environments, allowing the monitoring of the real-time results.

\begin{figure*}[htb]
\centering{} \includegraphics[scale=0.6]{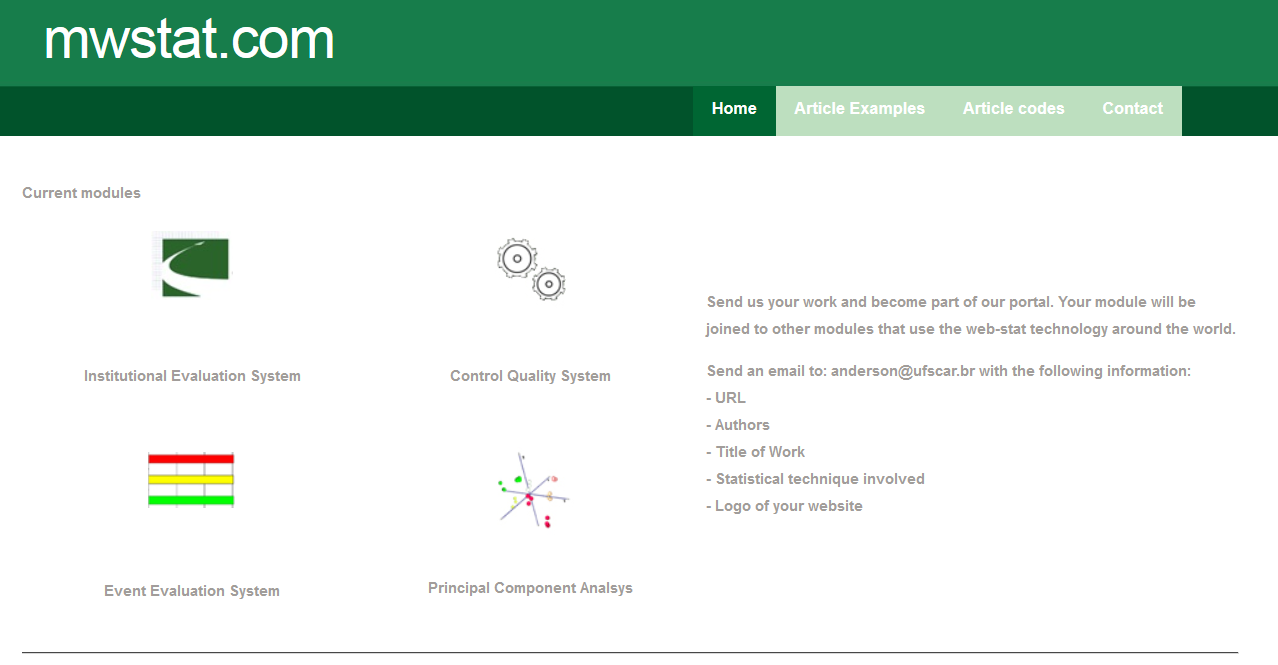} \caption{ Website of the MWStat where possible developers may provide their statistical analysis modules for general users. \label{fig:Site}}
\end{figure*}

\section{Final Comments}

\label{sec:Final-Comments}

This paper discuss a current technological trend based on statistical calculations directly online, and  displays the  necessary tools to build a module of the MWStat, a dynamic environment applied to online research. The MWStat results from the relationship of three programming languages (PHP,
MySQL and R), and it may be considered for performing many statistical tasks within a web page structure.
Our focus here is to display the necessary procedures so that the MWStat may be implemented and disseminated in other servers. The development of other aspects such as web forms is linked to the experience of the programmer who will use the server displayed here.

There is plenty of room for future developments of the MWStat. 
We believe almost all statistical techniques can be implemented by
using the strategy proposed in the paper shifting the statistical paradigm from offline to online analysis, accessible to anyone with only internet access. For instance, we envisage new MWStat modules on methodologies such as time series for financial online analysis, online statistical quality control, parametric and non-parametric bioequivalence tests, and so on. 
In general terms, by means of any method implemented on the R software, the MWStat can perform online collection, data analyses, results and even reports, providing real time up-to-date statistical methods available online to practitioners and researchers.

A possible challenge is to consider the MWStat for educational purposes, since it may provide a straightforward path for learning statistics, without the need of  installing any software but only an internet access.  

Figure \ref{fig:Site} presents the  MWStat homepage ( \url{http://www. mwstat.com}), where the presented examples may be tried as well as the entire codes of the first and second examples may be downloaded, providing an straightforward way of interested readers to reproduce the statistical analysis presented here in their own servers, as well as develop their own R web based statistical  applications. 

The MWStat has been used in several applied researchers in Brazil and abroad. We cite the development of the module SAO-Docentes (in Portuguese, SAO is the acronym for "Sistema de Avalia\c{c}\~{a}o Online" and Docentes denotes lectures), which was specifically built for providing a real time poll for teacher evaluation at the Federal University of S\~{a}o Carlos (Brazil).
During the years of 2010--2015, the SAO-Docente was answered for more than 10,000 students for teacher evaluation of more than 2,000 university courses.  
The module SAO-Egressos (in Portuguese, Egressos denotes graduates or egress students), which was built to provide information on an overall evaluation of the Federal University of S\~{a}o Carlos (Brazil) by its  egress students.
Over 2011--2014, more than 4,000 ex-students  were exposed to the SAO-Egressos. Some reports (in Portuguese) may be found in the site \url{www.cpa.ufscar.br}.
Further, the MWStat has been used for business satisfaction surveys. For instance, we cite a hotel poll conducted for a Brazilian hotel group in 2009 and 2010, a business poll conducted for a  Brazilian accounting company in 2009 and 2011, a question system for a Business and Commercial Chamber of two counties  in 2014.
The MWStat has also been used as part of the methodological structure of statistical analysis in masters and doctorate thesis, such as the "The national policy on technical assistance and rural extension: Perceptions and Trends" by \citet{kleber:10} and "Perception of the architectural work space for the university community: the UFSCar case study" by  \citet{beth:11}. 
Moreover, the MWStat has been successfully applied  for more than 40 congresses evaluations in Brazil, Portugal and Peru. We point out some of them such as the 54th Annual Meeting of the Brazilian Region of the International Biometrical Society and the 13th Symposium on Applied Statistics to Agronomic Experimentation, both held in S\~{a}o Carlos, Brazil, in 2009, the 3td School of Sampling and Research Methodology and the 2nd International Workshop on Surveys for Policy Evaluation, both held in Juiz de Fora, Brazil, in 2011, the 20th and 21st National Symposium of Probability and Statistics, held respectively in Jo\~{a}o Pessoa and Natal, Brazil, in 2012 and 2014, and the 60th ISI World Statistics Congress (WSC), held in Rio de Janeiro, Brazil, July 2015.

\bigskip \bigskip
\noindent \textbf{{\large \textbf{Acknowledgments}}}:  The researchers of Francisco Louzada and Anderson Ara are supported by the Brazilian organizations CNPq and FAPESP.

\bibliographystyle{elsarticle-harv}
\bibliography{myrefs}


\end{document}